\documentclass[11pt]{article}             
\usepackage{osid}                         %
\usepackage{amsmath}
\usepackage{amssymb}
\usepackage{amsmath,amsfonts,latexsym,graphicx,amssymb}
\newcommand{\Mn}{M_n(\mathbb{C})}
\newcommand{\Mq}{M_2(\mathbb{C})}
\newcommand{\Mk}{M_k(\mathbb{C})}
\newcommand{\Mm}{M_m(\mathbb{C})}
\newcommand{\Mnm}{M_{n \times m}(\mathbb{C})}
\newcommand{\Mnp}{M_n^+(\mathbb{C})}
\newcommand{\ra}{\,\rightarrow\,}

\newcommand{\ot}{{\,\otimes\,}}
\newcommand{{\Cd}}{{\mathbb{C}^d}}
\newcommand{{\Rn}}{{\mathbb{R}^n}}

\def\oper{{\mathchoice{\rm 1\mskip-4mu l}{\rm 1\mskip-4mu l}%
{\rm 1\mskip-4.5mu l}{\rm 1\mskip-5mu l}}}
\def\<{\langle}
\def\>{\rangle}
\newtheorem{thm}{Theorem}
\newtheorem{Problem}{Problem}
\newtheorem{cor}{Corollary}

\newtheorem{Definition}{Definition}
\newtheorem{Example}{Example}
\newtheorem{pro}{Proposition}
\newtheorem{remark}{Remark}

\title{\bf On time-local generators of quantum evolution\footnote{Lecture presented at the school {\em Open Quantum Systems}, Toru\'n, June 20-24, 2012; http://tv.umk.pl/\#movie=1743}}
\author{Dariusz Chru\'sci\'nski \\
Institute of Physics, Faculty of Physics, Astronomy and Informatics \\ Nicolaus Copernicus University \\
Grudziadzka 5, 87--100 Torun, Poland}

\begin{document}

\maketitle

\begin{abstract}
We present a basic introduction to the dynamics of open quantum systems based on local-in-time master equations. We characterize the properties of time-local generators giving rise to legitimate completely positive trace preserving quantum evolutions. The analysis of Markovian and non-Markovian quantum dynamics is presented as well. The whole discussion is illustrated by the family of many instructive examples.
\end{abstract}

\section{Introduction}

The dynamics of open quantum systems attracts nowadays considerable attention. It is relevant not only for a better understanding of
quantum theory but it is fundamental in various modern applications of quantum mechanics
such as quantum communication, cryptography, computation and quantum metrology. Any realistic quantum system is an {\em open system} since it always interacts with its environment or ``rest of the world". The proper descriptions of such systems is therefore of fundamental importance.  There are several excellent  books \cite{Breuer,Weiss,Davies,Alicki,Benatti-2003,Rivas} and review articles \cite{Benatti,Breuer-JPB} devoted to this subject. In this paper we provide an introduction to the dynamics of open quantum systems based on local-in-time master equation.
There are several approaches to the dynamics of open quantum systems. It is clear that each description presents a departure from the standard von Neumann equation
\begin{equation}\label{}
    i \dot{\rho}_t = [H,\rho_t]\ ,
\end{equation}
which governs the evolution of the {\em closed system} fully characterized by the system Hamiltonian $H$ (by $\rho_t$ we denote the density operator (density matrix) at time $t$ -- see the next section for the notation) and $\hbar=1$ throughout the paper. Introducing a linear map acting on the space of density operators (sometimes called a {\em superoperator}) $L(\rho) = -i[H,\rho]$ the von Neumann equation may be rewritten in a more compact way as $\dot{\rho}_t = L(\rho_t)$. $L$ is called a {\em generator} of the evolution. A natural way to depart from the world of {\em closed systems} is to change the corresponding equation of motion by changing the generator. The aim of this paper is to provide basic introduction to so called time-local master equation
\begin{equation}\label{M}
    \dot{\rho}_t = L_t(\rho_t)\ ,
\end{equation}
which  is fully characterized by the time-dependent generator $L_t$. Clearly, von Neumann equation is a special case of this general scheme.

There is an alternative approach (see e.g. \cite{Breuer}) which is based on the following non-local equation
\begin{equation}\label{NM}
    \dot{\rho}_t = \int_0^t \mathcal{K}_{t-\tau}(\rho_\tau) \, d\tau\ ,
\end{equation}
fully characterized by the so called {\em memory kernel} $\mathcal{K}_t$. Usually, the non-local character of (\ref{NM}) is attributed to the presence of quantum memory effects:   this simply means that
the rate of change of the state $\rho_t$ at time $t$ depends on its
history (starting at $t=0$). Note that time-local master equation (\ref{M}) with constant generator $L$ is reobtained when $\mathcal{K}_t = 2\delta(t)L$.

One of the fundamental problems in the theory of open quantum systems  is to find  conditions on $L_t$ and $\mathcal{K}_t$ that ensure that solutions to (\ref{M}) and (\ref{NM}) are physically legitimate. In the present paper we deal with conditions for time-local generator $L_t$ only. Surprisingly, it turns out that the problem of necessary and sufficient conditions for $L_t$ is open. Our aim is to show why this problem is difficult, what is already known and what are the perspectives.

The paper is organized as follows: in the next Section we introduce basic notation and recall basic notions we use in this paper like states, linear positive maps, quantum channels and all that. Section \ref{SEC-How} provides the description of quantum evolution in terms of time-local master equation. Sections \ref{SEC-C1}, \ref{SEC-C2} and  \ref{SEC-C3} characterize classes of local generators for which the corresponding conditions for $L_t$ may be easily formulated. We analyze time independent generators characterizing Markovian semigroups in Section  \ref{SEC-C1}, so called commutative dynamics in Section  \ref{SEC-C2}, and Markovian evolution in Section  \ref{SEC-C3}. Section \ref{SEC-GEN} provides a discussion on the general structure of time-local generator: we analyze a simple qubit dynamics to show an intricate structure of the corresponding local generator. Final conclusions are collected in the last section.

\section{Preliminaries: quantum states and quantum channels}

We begin by introducing basic notation and terminology.

\subsection{The structure of quantum states}

In this paper we consider a quantum system living in finite-dimensional Hilbert space $\mathcal{H}$ isomorphic to $\mathbb{C}^n$. Fixing an orthonormal basis $\{e_1,\ldots,e_n\}$ in $\mathcal{H}$ any linear operator in $\mathcal{H}$ may be identified with an $n \times n$ complex matrix, i.e. an element from $M_n(\mathbb{C})$.  A mixed state of such system is represented by a density matrix, i.e. a matrix  $\rho$ from $ M_n(\mathbb{C})$ such that $\,  \rho \geq 0$ and  ${\rm Tr}\, \rho=1$. A space of states $\mathfrak{S}_n$ of $n$-level quantum system defines an $(n^2-1)$-dimensional convex set.  Pure states correspond to rank-1 projectors $|\psi\>\<\psi|$ and define extremal elements of $\mathfrak{S}_n$. Any density matrix may be therefore decomposed as follows
\begin{equation}\label{}
    \rho  = \sum_k w_k |\psi_k\>\<\psi_k| \ ,
\end{equation}
with $w_k > 0$ and $\sum_k w_k =1$, i.e. $w_k$ provides a probability distribution. It should be stressed that the above decomposition is highly non-unique.

To illustrate a concept of density operators let us consider the following
\begin{Example} A 2-level system (qubit) living in $\mathbb{C}^2$. Any hermitian operator $\rho$ may be decomposed as follows
\begin{equation}\label{}
    \rho = \frac 12 ( \mathbb{I}_2 + \sum_{k=1}^3 x_k \sigma_k) \ ,
\end{equation}
where $\mathbf{x}=(x_1,x_2,x_3) \in \mathbb{R}^3$ and $\{\sigma_1,\sigma_2,\sigma_3\}$ are Pauli matrices. As usual $\mathbb{I}_n$ denotes a unit matrix in $M_n(\mathbb{C})$. It is, therefore, clear that $\rho$ is entirely characterized by the Bloch vector $\mathbf{x}$.
This representation already guaranties that ${\rm Tr}\, \rho=1$. Hence, $\rho$ represents density operator if and only if the corresponding eigenvalues $\{\lambda_-,\lambda_+\}$ are non-negative. One easily finds
\begin{equation}\label{}
    \lambda_- = \frac 12 (1 - |\mathbf{x}|) \ , \ \ \  \lambda_+ = \frac 12 (1 + |\mathbf{x}|) \ ,
\end{equation}
and hence $\rho \geq 0$ if and only if $|\mathbf{x}|=\sqrt{x_1^2 + x_2^2 + x_3^2} \leq 1$. This condition defines a unit ball in $\mathbb{R}^3$ known as a Bloch ball. A state is pure if $\rho$ defines rank-1 projector, i.e. $\lambda_-=0$ and $\lambda_+=1$. It shows that pure states belong to Bloch sphere corresponding to $|\mathbf{x}|=1$. Unfortunately, this simple geometric picture is much more complicated if $n >2$.
\end{Example}

For any $A \in \Mn$ we denote by $||A||_1 := {\rm Tr} |A|  = {\rm Tr}\sqrt{AA^\dagger}$ the trace-norm of $A$. If $\lambda_1,\ldots\lambda_n$ are (necessarily nonnegative) eigenvalues of $AA^\dagger$, then
$$   ||A||_1 = \sqrt{\lambda_1} + \ldots + \sqrt{\lambda_n} \ . $$
The space of states $\mathfrak{S}_n$ is equipped with a natural metric structure: given two states $\rho,\sigma \in \mathfrak{S}_n$ one defines the corresponding distance
\begin{equation}\label{}
    D[\rho,\sigma] = \frac 12 \,||\rho - \sigma||_1 \ .
\end{equation}
This quantity  measurers distinguishability between $\rho$ and $\sigma$. It is clear that $D[\rho,\sigma] =0$, i.e. $\rho$ and $\sigma$ are indistinguishable, if and only if $\rho=\sigma$. Note, that if $\rho$ and $\sigma$ are orthogonally supported, then
$$ D[\rho,\sigma] = \frac 12 ( ||\rho||_1 + ||\sigma||_1) = 1\ , $$
since $||\rho||_1=1$ for any density matrix $\rho$. In this case $\rho$ and $\sigma$ are perfectly distinguishable. Hence
\begin{equation}\label{}
    0 \leq D[\rho,\sigma] \leq 1\ .
\end{equation}

 In particular, if $\rho$ and $\sigma$ are two states of a qubit and $\mathbf{x}$ and $\mathbf{y}$ are corresponding Bloch vectors then
\begin{equation}\label{}
    D[\rho,\sigma] = \frac 12 \, |\mathbf{x} - \mathbf{y}|  \ ,
\end{equation}
reproduces standard Euclidean distance in $\mathbb{R}^3$.  For more information about the structure of quantum states see \cite{KAROL,FAZY}.

\subsection{Positive and completely positive maps \cite{Paulsen,Kadison,Bhatia}}

Consider now a linear map $\Phi : \Mn \ra \Mn$ and let  $\Mnp = \{ A \in \Mn\, |\, A\geq 0 \} \subset \Mn$ be a convex subset of positive matrices. One calls a linear map $\Phi$

\begin{itemize}

\item Hermicity-preserving  iff $\Phi(A^\dagger) = [\Phi(A)]^\dagger$,

\item positive iff $\Phi(\Mnp)\subset \Mnp$,

\item trace-preserving iff ${\rm tr}\, \Phi(A) = {\rm tr}\, A$,

\item unital iff $\Phi(\mathbb{I}_n) = \mathbb{I}_n$.

\end{itemize}

It is easy to show that positive map is necessarily Hermicity-preserving. Moreover, observing that $\mathfrak{S}_n = \{ A\in \Mnp\, |\ {\rm tr}\, A=1\}$ it is clear that if $\Phi$ is positive and trace preserving than it maps density matrices into density matrices, i.e. $\Phi(\mathfrak{S}_n) \subset \mathfrak{S}_n$. If $\Phi$ is a linear map then one defines a dual map $\Phi^* : \Mn \ra \Mn$ by
\begin{equation}\label{}
    {\rm Tr}[A \Phi^*(B)] = {\rm Tr}[\Phi(A)B]\ ,
\end{equation}
for all $A,B\in \Mn$. $\Phi$ is trace-preserving iff $\Phi^*$ is unital.

\begin{Example}\label{EX-2} Consider a transposition $T_n : M_n(\mathbb{C})\ra \Mn$, i.e. $T_n(\rho) = \rho^T$. Since transposition does not affect the spectrum of $A$ it clear that $A^T \geq 0$ whenever $A\geq 0$. Note that $T_n$ is trace-preserving and unital. As another example consider $R_n : \Mn \ra \Mn$ defined by
\begin{equation}\label{}
    R_n(A) = \frac{1}{n-1} ( \mathbb{I}_n\, {\rm Tr}\, A - A)\ .
\end{equation}
One calls $R_n$ a {\em reduction map}. Let $|x\> \in \mathbb{C}^n$ and $P_x :=|x\>\<x|$ be the corresponding  rank-1 projector (we assume that $\<x|x\>=1$). One finds $R_n(P_x) = \frac{1}{n-1} P_x^\perp$, where $P_x^\perp$ is a projector complementary to $P_x$, i.e. $P_x + P_x^\perp = \mathbb{I}_n$. Since any $A\geq 0$ is a convex combination of rank-1 projectors it proves the positivity of $R_n$. Moreover, $R_n$ is trace-preserving and unital.
\end{Example}

\begin{remark} Note that fixing an orthonormal basis $\{e_k\}$ in $\mathcal{H}$ and defining $P_k = |e_k\>\<e_k|$  one easily shows that if $\Phi$ is a positive trace-preserving map, then the following $n \times n$ matrix
\begin{equation}\label{}
    T_{ij} = {\rm Tr}( P_i \Phi(P_j) ) \ ,
\end{equation}
is stochastic. Recall, that $A \in M_n(\mathbb{R})$ is a stochastic matrix iff  $A_{ij} \geq 0$ and $\sum_{i=1}^n A_{ij} =1$ for all $j=1,\ldots,n$.  Indeed, $T_{ij} \geq 0$ and
\begin{equation}\label{}
    \sum_{i=1}^n  T_{ij} = {\rm Tr}\Big( \Big[ \sum_{i=1}^n P_i \Big] \Phi(P_j) \Big) = {\rm Tr}\, [\Phi(P_j)] =1 \ ,
\end{equation}
for all $j=1,\ldots,n$. It is clear, that if $\Phi$ is also unital, then $T_{ij}$ is doubly stochastic.
\end{remark}
Positive trace-preserving maps possess the following fundamental property

\begin{pro}[\cite{Paulsen,Kadison,Bhatia}] \label{TH-D}If $\Phi$ is positive and trace-preserving, then
\begin{equation}\label{}
    || \Phi(X)||_1 \leq ||X||_1\ ,
\end{equation}
for all $X \in \Mn$, that is, $\Phi$ is a contraction in trace-norm. Hence
\begin{equation}\label{}
    D[\Phi(\rho),\Phi(\sigma)] \leq D[\rho,\sigma]\ ,
\end{equation}
which means that distinguishability of $\rho$ and $\sigma$ never increases under the action of positive and trace-preserving map.
\end{pro}
It turns out that positivity property is not sufficient for quantum physics. It is connected with the notion of composed systems. Composing two systems living in  $\mathcal{H}_1$ and $\mathcal{H}_2$, respectively, one obtains a system living in $\mathcal{H}=\mathcal{H}_1 \ot \mathcal{H}_2$. Let ${\rm dim}\, \mathcal{H}_1 = n$,  ${\rm dim}\, \mathcal{H}_2 = m$ and consider two linear maps
$$ \Phi_1 : \Mn \ra  \Mn\ , \ \ \ \ \Phi_2 : \Mm \ra  \Mm\ \ . $$
Recalling that $\Mnm = \Mn \ot \Mn$ one defines a tensor product
$$ \Phi_1 \ot \Phi_2 : \Mnm \ra  \Mnm\ ,$$
as follows: for a fixed  orthonormal basis $\{e_1,\ldots,e_n\}$ in $\mathcal{H}_1$ let us define $e_{ij} := |e_i\>\<e_j| \in \Mn$. Elements $\{e_{ij}\}$ for $i,j=1,\ldots,n$ define an orthonormal basis in $\Mn$ with respect to the standard inner product $(A,B) = {\rm tr}(A^\dagger B)$. Now, any matrix $A \in \Mnm$ may be represented in the following block form
\begin{equation}\label{}
    A = \sum_{i,j=1}^n e_{ij} \ot A_{ij}\ ,
\end{equation}
with $A_{ij} \in \Mm$. For example if $n=2$ one has
\begin{equation}\label{}
    A = \sum_{i,j=1}^2 e_{ij} \ot A_{ij} = \left( \begin{array}{c|c} A_{11} & A_{12} \\ \hline A_{21} & A_{22} \end{array} \right) \ .
\end{equation}
Hence the action of $\Phi_1 \ot \Phi_2$ is given by
\begin{equation}\label{}
    [\Phi_1 \ot \Phi_2](A) := \sum_{i,j=1}^n \Phi_1(e_{ij}) \ot \Phi_2(A_{ij})\ .
\end{equation}
In particular if $n=2$ and $\Phi_1 = \oper_2$, where $\oper_n : \Mn \ra \Mn$ denotes an identity map defined by $\oper_n(X) = X$, then
\begin{equation}\label{}
   [\oper_2 \ot \Phi](A) = \sum_{i,j=1}^2 e_{ij} \ot \Phi(A_{ij}) = \left( \begin{array}{c|c} \Phi(A_{11}) & \Phi(A_{12}) \\ \hline \Phi(A_{21}) & \Phi(A_{22}) \end{array} \right) \ .
\end{equation}
Now comes a surprise: even if $\Phi_1$ and $\Phi_2$ are positive $\Phi_1 \ot \Phi_2$ needs not be a positive map.

\begin{Example} Interestingly, both maps considered in Example \ref{EX-2} loose their positivity when tensoring with other positive maps. This map is evidently positive and trace-preserving. Consider $\oper_2 \ot T_2$ and $\oper_2 \ot R_2$. It turns out that these maps are not positive in $M_4(\mathbb{C})$. Indeed, let
\begin{equation}\label{}
     P^+_2 = \frac 12 \sum_{i,j=1}^2 e_{ij} \ot e_{ij} = \frac 12 \left( \begin{array}{cc|cc} 1 & 0 & 0 & 1 \\ 0 & 0 & 0 & 0 \\ \hline   0 & 0 & 0 & 0 \\ 1 & 0 & 0 & 1 \end{array} \right) \ ,
\end{equation}
be a state in $\mathbb{C}^2 \ot \mathbb{C}^2$. Note that $P^+_2 = |\psi^+_2\>\<\psi^+_2|$ with $\psi^+_2 = (e_1 \ot e_1 + e_2 \ot e_2)/\sqrt{2}$ being one of the well-known Bell states of two qubits. One finds
\begin{equation}\label{}
    [\oper_2 \ot T_2]( P^+_2) = \frac 12 \sum_{i,j=1}^2 e_{ij} \ot e_{ji} = \frac 12 \left( \begin{array}{cc|cc} 1 & 0 & 0 & 0 \\ 0 & 0 & 1 & 0 \\ \hline   0 & 1 & 0 & 0 \\ 0 & 0 & 0 & 1 \end{array} \right) \ ,
\end{equation}
and
\begin{equation}\label{}
    [\oper_2 \ot R_2]( P^+_2) = \frac 12 \sum_{i,j=1}^2 e_{ij} \ot R_2(e_{ij}) = \frac 12 \left( \begin{array}{cc|cc} 0 & 0 & 0 & -1 \\ 0 & 1 & 0 & 0 \\ \hline   0 & 0 & 1 & 0 \\ -1 & 0 & 0 & 0 \end{array} \right) \ .
\end{equation}
Note that both matrices $[\oper_2 \ot T_2]( P^+_2)$ and $[\oper_2 \ot R_2]( P^+_2)$ have one negative eigenvalue and hence neither $\oper_2 \ot T_2$ nor $\oper_2 \ot R_2$ is a positive map.
\end{Example}
This example proves that  quantum physics of composed systems  needs a more refined notion of positivity.  Consider again a linear map $\Phi : \Mn \ra \Mn$. One calls $\Phi$ $k$-positive if
\begin{equation}\label{}
\oper_k \ot \Phi : \Mk \ot \Mn \ra \Mk \ot \Mn \ ,
\end{equation}
is positive. Clearly 1-positive is just positive and $k$-positivity implies $\ell$-positivity for $\ell < k$. Finally, $\Phi$ is called {\em completely positive} (CP) if it is $k$-positive for $k=1,2,\ldots$.  Interestingly, one has the following characterization

\begin{pro}[Choi \cite{Choi}]
If ${\rm dim}\, \mathcal{H} = n$, then $\Phi$ is CP if and only if $\Phi$ is $n$-positive.
\end{pro}
Denoting by $\mathcal{P}_k$ a convex set of $k$-positive maps one has the following chain of inclusions
\begin{equation*}\label{}
    {\rm CP} \equiv \mathcal{P}_n \subset \ldots \subset \mathcal{P}_2 \subset \mathcal{P}_1 \equiv {\rm Positive\ maps} \ .
\end{equation*}
Let $\{e_1,\ldots,e_n\}$ be a fixed orthonormal basis in $\mathcal{H}$ and let
\begin{equation}\label{}
 |\psi^+_n\> = \frac{1}{\sqrt{n}}\, \sum_{k=1}^n e_k \ot e_k\ ,
\end{equation}
denote a maximally entangled state in $\mathcal{H} \ot \mathcal{H}$. Moreover,  let $P^+_n = |\psi^+_n\>\<\psi^+_n|$ denote the corresponding rank-1 projector.
\begin{pro}[Choi \cite{Choi}]
$\Phi$ is CP if and only if $[\oper_n \ot \Phi](P^+_n) \geq 0$.
\end{pro}
This beautiful result states that in order to prove that $\Phi$ is CP, or equivalently that $\oper_n \ot \Phi$ is positive, it is enough to check wether $\oper_n \ot \Phi$ is positive on one particular projector $P^+_n$. Positivity of $[\oper_n \ot \Phi](P^+_n)$ guaranties that $[\oper_n \ot \Phi](X) \geq 0$ for all positive $X \in \Mn \ot \Mn$.

\begin{cor} If $\Phi_1$ and $\Phi_2$ are CP maps, then $\Phi_1 \ot \Phi_2$ is always CP as well.
\end{cor}
This analysis shows that the motivation to use CP maps is deeply rooted in physics and it is not just mathematical trick! It is the very presence of quantum entangled states enforces us to deal with maps which are completely positive. The following result provides the most important characterization of CP maps.
\begin{thm}[\cite{Paulsen,Kadison,Choi,Kraus}]
A map $\Phi : \Mn \ra \Mn$ is CP if and only if
\begin{equation}\label{Kraus}
\Phi(X) = \sum_\alpha\, K_\alpha X\, K_\alpha^\dagger\ ,
\end{equation}
for $X \in M_n(\mathbb{C})$.
\end{thm}
Formula (\ref{Kraus}) is usually called  Kraus or {\em Operator Sum Representation} of $\Phi$ and $K_\alpha$ are called Kraus operators. Actually, the above formula  appeared already in the Sudarshan et. al. paper \cite{Sudarshan}. It should be stressed that the Kraus representation is highly non unique. Completely positive trace preserving map (CPTP) is called a {\em quantum channel}. A CP map possessing Kraus representation is trace-preserving iff
\begin{equation}\label{Kraus-KIK}
 \sum_\alpha\, K_\alpha^\dagger \, K_\alpha = \mathbb{I}_n\ .
\end{equation}
The following result shows what is the origin of a genuine quantum channel.

\begin{thm}[Unitary dilation] Any quantum channel $\Phi$ may be constructed as follows
\begin{equation}\label{dilation}
    \Phi(\rho) = {\rm tr}_E \big[ U (\rho \ot \omega) U^\dagger \big]\ ,
\end{equation}
where $U$ is a unitary operator in $\mathcal{H}\ot \mathcal{H}_E$ and $\omega$ is a density operator in $\mathcal{H}_E$, and ${\rm Tr}_E$ denotes partial trace over $\mathcal{H}_E$.
\end{thm}
One usually interprets $\mathcal{H}_E$ as a Hilbert space of the environment and $\omega$ as its fixed state.
Let
$$ \omega |E_k\>  =  \lambda_k |E_k\> \  , $$
with $\lambda_k \geq 0$.  Moreover,  let $U = \sum_{k,l} U_{kl} \ot |E_k\>\<E_l|$. Formula (\ref{dilation}) implies
\begin{eqnarray*}
    \Phi(\rho) &=& \sum_{m,n} \sum_{i,j} \sum_k \lambda_k \, {\rm Tr}_E \big[ (U_{ij} \ot |E_i\>\<E_j|)  (\rho \ot |E_k\>\<E_k|) (U_{mn}^\dagger \ot |E_n\>\<E_m|) \big] \nonumber \\ &=& \sum_{m,n} \sum_{i,j} \sum_k \lambda_k \, {\rm Tr} [ |E_i\>\<E_j|E_k\>\<E_k|E_n\>\<E_m|]\, U_{ij} \rho U_{mn}^\dagger \ .
\end{eqnarray*}
Using ${\rm Tr} [ |E_i\>\<E_j|E_k\>\<E_k|E_n\>\<E_m| ] = \delta_{im} \delta_{jk} \delta_{kn}$ and introducing $K_\alpha := K_{mn} = \sqrt{\lambda_n}\, U_{mn}$ one arrives at the Kraus representation $\Phi(\rho) = \sum_\alpha\, K_\alpha \rho\, K_\alpha^\dagger$ which proves that $\Phi$ defined {\em via} formula (\ref{dilation}) is completely positive. One easily proves that $\Phi$ is also trace preserving and hence defines a quantum channel.

\section{How to describe quantum evolution}   \label{SEC-How}

If $\rho$ is an initial state of $n$-level quantum system, then by its evolution we mean a trajectory $\rho_t$ for $t \geq 0$ starting at $\rho$. The simplest example of quantum evolution is provided by the von Neumann equation
\begin{equation}\label{v-N}
    i\dot{\rho}_t = [H,\rho_t]\ ,
\end{equation}
with the corresponding solution
\begin{equation}\label{}
    \rho_t = \mathcal{U}_t (\rho)\ ,
\end{equation}
where the map $\mathcal{U}_t : \Mn \ra \Mn$ is defined by
\begin{equation}\label{UUU}
    \mathcal{U}_t(\rho) := U_t \rho U_t^\dagger\ ,
\end{equation}
with $U_t = e^{-iHt}$. Note that (\ref{UUU}) defines a family of quantum channels. Let us observe that 1-parameter unitary group $U_t$ implies the following composition law
\begin{equation}\label{U-t-s}
    \mathcal{U}_t \, \mathcal{U}_s = \mathcal{U}_{t+s}\ ,
\end{equation}
for all $t,s \in \mathbb{R}$. Hence $\mathcal{U}_t$ defines 1-parameter group of CP maps.
Equation (\ref{v-N}) may be transformed into the following equation for $\mathcal{U}_t$
\begin{equation}\label{v-N-U}
    \dot{\mathcal{U}}_t  = L \,\mathcal{U}_t\ , \ \ \ \mathcal{U}_0 = \oper\ ,
\end{equation}
where the generator $L : \Mn \ra \Mn$ is defined by
\begin{equation}\label{}
L(X) = -i [H,X]\ ,
\end{equation}
for any $X \in \Mn$. Now come  natural questions:

\begin{enumerate}

\item how to generalize the unitary evolution defined by (\ref{UUU}), and

\item how to generalize the corresponding equation of motion (\ref{v-N-U})?

\end{enumerate}

\begin{Definition}
By a general quantum evolution we mean a dynamical map, i.e. a family of quantum channels $\Lambda_t : \Mn \ra \Mn$ for $t \geq 0$ such that $\Lambda_0 = \oper_n$.
\end{Definition}
A dynamical map $\Lambda_t$ maps an initial state $\rho$ into a current state $\rho_t := \Lambda_t(\rho)$ and hence provides natural generalization of the unitary evolution $\rho_t = \mathcal{U}_t(\rho)$. Assuming that $\rho_t$ satisfies a linear equation and that the initial state $\rho$ provides all necessary information to uniquely determine $\rho_t$ we expect that $\rho_t$ satisfies the following equation
\begin{equation}\label{}
    \dot{\rho}_t = L_t(\rho_t)\ ,
\end{equation}
or equivalently
\begin{equation}\label{ME}
    \dot{\Lambda}_t = L_t\Lambda_t\ ,\ \ \ \Lambda_0 = \oper_n\ ,
\end{equation}
where $L_t : \Mn \ra \Mn$ denotes time-dependent generator. This equation provides a natural generalization of (\ref{v-N-U}).
The formal solution of (\ref{ME}) may be written as follows
\begin{equation}\label{Lambda-t}
    \Lambda_t = {\rm T} \exp\left( \int_0^t L_\tau\, d\tau\right)\ ,
\end{equation}
where T denotes the chronological product. The above formula is defined by the following Dyson series
\begin{equation}\label{Dyson}
    \Lambda_t = \oper_n + \int_0^t dt_1\,  L_{t_1} + \int_0^{t} dt_1 \int_0^{t_1} dt_2 \,  L_{t_1} \,  L_{t_2} + \ldots \ ,
\end{equation}
provided that it converges. In this paper we address the following

\begin{Problem}
What are the properties of local time-dependent generator $L_t$ which guarantee that $\Lambda_t$ defined by T-product exponential formula (\ref{Lambda-t}) defines a legitimate dynamical map?
\end{Problem}
The formulation of our problem is pretty simple, however, in general the answer is not known.
Let us observe that if we knew a dynamical map $\Lambda_t$ which was invertible, i.e. there exists $\Lambda_t^{-1} : \Mn \ra \Mn$ such that $\Lambda_t^{-1} \Lambda_t = \Lambda_t \Lambda_t^{-1} = \oper_n$, then
\begin{equation}\label{}
    \dot{\Lambda}_t = \dot{\Lambda}_t \Lambda_t^{-1} \Lambda_t = L_t \, \Lambda_t\ ,
\end{equation}
where we defined
\begin{equation}\label{}
    L_t := \dot{\Lambda}_t \Lambda_t^{-1} \ .
\end{equation}
It should be stressed that the inverse of $\Lambda_t$ needs not be CP. One may prove that if $\Lambda_t$ is CP then $\Lambda_t^{-1}$ is CP  if and only if $\Lambda_t(\rho) = U_t \rho U_t^\dagger$ with unitary $U_t$.

\begin{Example} Consider a unitary dynamical map $\mathcal{U}_t$ defined in (\ref{UUU}). It is clear that $\mathcal{U}_t$ is invertible and $\mathcal{U}_t^{-1} = \mathcal{U}_{-t}$ is CP. One finds for the corresponding generator
\begin{equation}\label{}
    L_t(\rho) = [\dot{\mathcal{U}}_t \mathcal{U}_{-t}](\rho)  = \dot{U}_t (U_t^\dagger \rho U_t) U_t^\dagger + {U}_t (U_t^\dagger \rho U_t) \dot{U}_t^\dagger \ ,
\end{equation}
and hence recalling that $U_t$ satisfies Schr\"odinger equation $\dot{U}_t = -iHU_t$, one obtains $L_t(\rho) = -i[H,\rho]$.
\end{Example}

In this paper instead of analyzing this problem in full generality we restrict ourselves to study special classes of dynamical maps and corresponding local generators.
In what follows we analyze 3 important classes of generators

\begin{itemize}

\item $C_1$ --- a class of time independent generators giving rise to Markovian semigroup,

\item $C_2 $ --- a class of time dependent generators giving rise to commutative dynamics

\item $C_3 $ --- a class of time dependent generators giving rise to the so-called divisible dynamical maps.

\end{itemize}
Our analysis proves the following relations between these classes: $\   C_1 \subset C_2 \cap C_3$.

\section{Markovian semigroups}   \label{SEC-C1}

In this section we recall the celebrated result derived independently by Gorini, Kossakowski and Sudarshan \cite{GKS} and Lindblad \cite{Lindblad}.
Consider the simplest case of a time independent generator $L$
\begin{equation}\label{ME0}
    \dot{\Lambda}_t = L\Lambda_t\ ,\ \ \ \Lambda_0 = \oper_n\ .
\end{equation}
The formal solution is given by $\Lambda_t = e^{tL}$ for $t \geq 0$ and hence satisfies the following composition law
\begin{equation}\label{c-low}
    \Lambda_t \, \Lambda_s = \Lambda_{t+s}\ ,
\end{equation}
for $t,s \geq 0$. It proves that $\Lambda_t$ provides a semigroup of linear maps (it is a semigroup since the inverse $\Lambda_t^{-1}$ needs not be  CP). The properties of $L$ are summarized in the following

\begin{thm}[\cite{GKS,Lindblad}] \label{TH-GKSL}
A linear map $L : \Mn \ra \Mn$ generates legitimate dynamical semigroup if and only if
\begin{equation}\label{MAIN}
    L(\rho) = -i[H,\rho] + \Phi(\rho) - \frac 12 \{ \Phi^*(\mathbb{I}_n),\rho\} \ ,
\end{equation}
where $\Phi : \Mn \ra \Mn$ is CP, $\Phi^*$ denotes the dual map, and $H=H^\dagger \in \Mn$.
\end{thm}
In what follows we call $L$ satisfying (\ref{MAIN}) a GKSL generator.
\begin{remark} If $\Phi$ is only a positive map, then $L$ generate a semigroup of positive maps $\Lambda_t$ which needs not be CP.
\end{remark}
Let
\begin{equation}\label{kr}
    \Phi(\rho) = \sum_k V_k \rho V_k^\dagger\ ,
\end{equation}
be a Kraus representation of $\Phi$. Its dual is represented by $ \Phi^*(X) = \sum_k V_k^\dagger X V_k$.

\begin{cor} A generator of a dynamical semigroup can be written in the following form
\begin{equation}\label{}
    L(\rho) = -i[H,\rho] +  \sum_k \left( V_k\rho V_k^\dagger - \frac 12 \{   V_k^\dagger V_k, \rho\} \right) \ ,
\end{equation}
or equivalently
\begin{equation}\label{}
    L(\rho) = -i[H,\rho] + \frac 12 \sum_k \left( [V_k,\rho V_k^\dagger] + [V_k\rho,V_k^\dagger] \right) \ .
\end{equation}
\end{cor}

\begin{remark} Note, that if $\Phi$ is CP and trace preserving (i.e. a quantum channel), then $\Phi^*(\mathbb{I}_n) = \mathbb{I}_n$ and hence
the formula (\ref{MAIN}) simplifies to
\begin{equation}\label{MAIN1}
    L(\rho) = -i[H,\rho] + \Phi(\rho) - \rho\ .
\end{equation}
\end{remark}
Note that the dual map $L^*$ defines the generator of  quantum evolution in the Heisenberg picture
\begin{equation}\label{}
    L^*(X) = i[H,X] + \Phi^*(X) - \frac 12 \{ \Phi^*(\mathbb{I}_n),X\} \ ,
\end{equation}
or using Kraus representation (\ref{kr})
\begin{equation}\label{}
    L^*(X) = i[H,X] + \frac 12 \sum_k \left( [V^\dagger_k,X V_k] + [V_k^\dagger X,V_k] \right) \ .
\end{equation}

Let us consider simple examples of Markovian semigroups $\Lambda_t$ and corresponding generators $L$.

\begin{Example} Suppose that $\Phi : \Mn \ra \Mn$ is a quantum channel such that $\Phi$ is a CP projector, i.e. $\Phi^2=\Phi$,  and consider
\begin{equation}\label{L-Phi}
    L = \gamma(\Phi - \oper_n)\ ,
\end{equation}
where $\gamma >0$ and we assumed $H=0$. One finds  the corresponding solution
\begin{eqnarray}\label{}
    \Lambda_t &=& e^{tL} = e^{-\gamma t} e^{\gamma t\Phi} = e^{-\gamma t} \left(  \oper_n + \gamma t \Phi  + \frac 12 (\gamma t)^2 \Phi + \ldots  \right)  \nonumber \\
    &=&  e^{-\gamma t} \oper_n + (1-e^{-\gamma t}) \Phi\ ,
\end{eqnarray}
that is, $\Lambda_t$ is a convex combination of two quantum channels: $\oper_n$ and $\Phi$. A typical example of a completely positive projector is given by
\begin{equation}\label{}
    \Phi(\rho) = \sum_{k=1}^n P_k \rho P_{k}\ ,
\end{equation}
where $P_k = |k\>\<k|$, i.e. $\Phi$ projects $\rho$ onto the diagonal: $\Phi(\rho) = \sum_k \rho_{kk} P_k$.  Hence
\begin{eqnarray}\label{}
    \rho_t = e^{-\gamma t} \rho  + (1-e^{-\gamma t}) \sum_{k=1}^n P_k \rho\, P_{k} \ ,
\end{eqnarray}
which shows that diagonal elements $\rho_{kk}$ remain invariant and off-diagonal $\rho_{kl}$ are multiplied by the damping factor $e^{-\gamma t}$. It is, therefore, clear that this dynamics describes pure decoherence with respect to an orthonormal basis $\{|1\>,\ldots,|n\>\}$.

\end{Example}

\begin{Example} Suppose that $\Phi : \Mn \ra \Mn$ is a quantum channel such that $\Phi^2=\oper_n$ and consider $L$ defined by (\ref{L-Phi}).
 One finds  the corresponding solution
\begin{eqnarray}\label{}
    \Lambda_t &=& e^{tL} = e^{-\gamma t} e^{\gamma t \Phi} = e^{-\gamma t} \left(  \oper_n + \gamma t \Phi  + \frac 12 (\gamma t)^2 \oper_n  + \frac{1}{3!} (\gamma t)^3 \Phi + \ldots  \right)  \nonumber \\
    &=&  e^{-\gamma t} \Big[ \cosh(\gamma t) \oper_n + \sinh(\gamma t) \Phi \Big]\ ,
\end{eqnarray}
or equivalently
\begin{equation}\label{}
    \Lambda_t = \frac 12 ( 1+ e^{-2\gamma t}) \oper_n + \frac 12 (1- e^{-2\gamma t}) \Phi\ ,
\end{equation}
which is another convex combination of $\oper_n$ and $\Phi$. To illustrate this class consider $n=2$ and let $\Phi(\rho) = \sigma_z \rho \sigma_z$ which gives rise to the following generator
\begin{equation}\label{}
    L(\rho) = \gamma ( \sigma_z \rho \sigma_z - \rho)\ .
\end{equation}
One finds $\, \Phi^2(\rho) = \sigma_z(\sigma_z \rho \sigma_z)\sigma_z = \rho$ and hence the corresponding evolution is given by
\begin{equation}\label{}
    \rho_t = \Lambda_t(\rho) = \frac 12 ( 1+ e^{-2\gamma t})\, \rho + \frac 12 ( 1- e^{-2\gamma t})\, \sigma_z \rho \sigma_z\ .
\end{equation}
Again this corresponds to a pure decoherence: $\rho_{12} \ra e^{-2 \gamma t}\rho_{12}$ while the diagonal elements $\rho_{11}$ and $\rho_{22}$ remain invariant.
\end{Example}

\begin{Example} Let us consider a qubit generator defined by $H = \frac \omega 2 \sigma_z$ and the following CP map
\begin{equation}\label{}
    \Phi(\rho) = \gamma_1 \sigma_+ \rho\, \sigma_+^\dagger + \gamma_2 \sigma_-  \rho\,\sigma_-^\dagger + \gamma \sigma_z  \rho\, \sigma_z\ ,
\end{equation}
where $\sigma_+ = |2\>\<1|$ and $\sigma_-=|1\>\<2|=\sigma_+^\dagger$ are standard qubit raising and lowering operators. The corresponding generator reads $L(\rho) = - i[H,\rho] + L_D(\rho)$ with the dissipative part
\begin{equation}\label{}
L_D = \frac{\gamma_1}{2}  L_1 + \frac{\gamma_2}{2}  L_2 + \frac{\gamma}{2} L_z\ ,
\end{equation}
where
\begin{eqnarray} \label{LLL}
    L_1(\rho) &=& [\sigma_+, \rho\sigma_-] + [\sigma_+ \rho, \sigma_-]  \ ,\nonumber \\
    L_2(\rho) &=& [\sigma_-, \rho\sigma_+] + [\sigma_- \rho,\sigma_+] \ , \\
    L_3(\rho) &=& \sigma_z \rho\sigma_z - \rho \ . \nonumber
\end{eqnarray}
$L_1$ corresponds to pumping (heating) process, $L_2$ corresponds to relaxation (cooling), and $L_3$ is responsible for  pure decoherence.
To solve the master equation $\dot{\rho}_t = L\rho_t$ let us parameterize $\rho_t$ as follows
\begin{equation}\label{}
    \rho_t = p_1(t) P_1 + p_2(t) P_2 + \alpha(t) \sigma_+ + \overline{\alpha(t)} \sigma_-\ ,
\end{equation}
with $P_k=|k\>\<k|$. Using the following relations
\begin{eqnarray*}
  L(P_1) &=& {\gamma_1} (P_2 - P_1) = - \gamma_1\, \sigma_3\ , \\
  L(P_2) &=& {\gamma_2} (P_1 - P_2) = \gamma_2\, \sigma_3 \ , \\
  L(\sigma_+) &=& (i \omega - \eta)\, \sigma_+\ ,\\
  L(\sigma_-) &=& (-i\omega - \eta)\, \sigma_-\ ,
\end{eqnarray*}
where
 $   \eta = \frac{\gamma_1 +\gamma_2}{2} + \gamma$,
one finds the following Pauli master equations  for the probability distribution $(p_1(t),p_2(t))$
\begin{eqnarray}
  \dot{p}_1(t)  &=  &- \gamma_1\, p_1(t) + \gamma_2\, p_2(t) \ , \\
  \dot{p}_2(t)  &=  & \ \ \, \gamma_1\, p_1(t) - \gamma_2\, p_2(t) \ ,
\end{eqnarray}
together with $\alpha(t) = e^{(i \omega - \eta)t}\alpha(0)$.
The corresponding solution reads
\begin{eqnarray}
  p_1(t) &=& p_1(0)\, e^{-( \gamma_1 + \gamma_2)t} + p_1^* \Big[ 1 - e^{-( \gamma_1 + \gamma_1)t}  \Big] \ ,\\
  p_2(t) &=& p_2(0)\, e^{-( \gamma_1 + \gamma_2)t} + p_2^* \Big[ 1 - e^{-( \gamma_2 + \gamma_2)t}  \Big] \ ,
\end{eqnarray}
where we introduced
\begin{equation}\label{p*}
    p_1^* = \frac{\gamma_1}{\gamma_1 + \gamma_2}\ , \ \ \ \   p_2^* = \frac{\gamma_2}{\gamma_1 + \gamma_2}\ .
\end{equation}
Hence, we have purely classical evolution of probability vector $(p_1(t),p_2(t))$ on the diagonal of $\rho_t$ and very simple evolution of the off-diagonal element $\alpha(t)$. Note, that asymptotically one obtains completely decohered density operator
$$  \rho_t \ \ \longrightarrow \ \ \left( \begin{array}{cc} p_1^* & 0 \\ 0 & p_2^* \end{array} \right) \ . $$
In particular if $\gamma_1=\gamma_2$ a state $\rho_t$ relaxes to the maximally mixed state (a state becomes completely depolarized).

\end{Example}

\section{Commutative dynamics}   \label{SEC-C2}

We call a dynamical map $\Lambda_t$ commutative if $[\Lambda_t,\Lambda_u]=0$ for all $t,u\geq 0$. It means that for each $A\in \Mn$ one has
\begin{equation}\label{}
    \Lambda_t(\Lambda_u(A)) = \Lambda_u(\Lambda_t(A)) \ .
\end{equation}
It is easy to show that commutativity of $\Lambda_t$ is equivalent to commutativity of the local generator
\begin{equation}\label{[]=0}
    [L_t,L_u]=0\ ,
\end{equation}
for any $t,u \geq 0$.  Note that in this case  the formula (\ref{Dyson}) considerably simplifies: the `T' product drops out and the solution is fully controlled by the integral $\int_0^t L_u du$:
\begin{equation}\label{Dyson-com}
    \Lambda_t = \exp\left( \int_0^t L_u du\right) = \oper_n + \int_0^t   L_{u}du + \frac 12 \left( \int_0^{t}  L_{u}du \right)^2 + \ldots \ .
\end{equation}
Now, it follows from Theorem \ref{TH-GKSL} that if $\Lambda = e^M$, then $\Lambda$ is a quantum channel if $M$ is GKSL generator.
Therefore, one has the following

\begin{thm} If $L_t$ satisfies (\ref{[]=0}), then $L_t$ is a legitimate generator if $\int_0^t L_\tau d\tau$ is a GKSL generator for all $t\geq 0$.
\end{thm}
Note, that if $L_t = L$ is time independent, then $\int_0^t L_udu = tL$ and the above theorem reproduces Theorem \ref{TH-GKSL}.
It is clear that if $L$ is a legitimate GKSL generator and $f: \mathbb{R}_+ \ra \mathbb{R}$ an arbitrary function, then $L_t = f(t) L$ generates a commutative dynamical map $\Lambda_t$ iff $\int_0^t f(u) du \geq 0$ for all $t\geq 0$. A typical example of commutative dynamics is provided by
\begin{equation}\label{}
    L_t = \omega(t) L_0 + a_1(t) L_1 + \ldots + a_N(t) L_N\ ,
\end{equation}
where $[L_\alpha,L_\beta]=0$ with $L_0(\rho) =  -i[H,\rho]$, and for $\alpha>0$ the generators $L_\alpha$ are purely dissipative, that is, $L_\alpha(\rho) = \Phi_\alpha(\rho) - \frac 12 \{ \Phi_\alpha^*(\mathbb{I}),\rho\}$.  One has for the corresponding dynamical map
\begin{equation}\label{}
    \Lambda_t = e^{\Omega(t)L_0} \cdot e^{A_1(t)L_1} \cdot \ldots\cdot e^{A_N(t)L_N}\ ,
\end{equation}
with
$$ \Omega(t) = \int_0^t \omega(u)du\ ; \ \ \ \ A_\alpha(t) = \int_0^t a_\alpha(u)du\  . $$
It is clear that $\Lambda_t$ is CP iff $A_\alpha(t) \geq 0$ for all $\alpha=1,\ldots,N$.

\begin{Example}
Consider qubit generator $L_0(\rho) = -i[\sigma_3,\rho]$ together with $L_1,L_2,L_3$ defined in (\ref{LLL}). One easily proves
\begin{equation}\label{}
    [L_0,L_\alpha] = [L_3,L_\alpha]=0 \ ; \ \ \ \alpha=1,2,3\ ,
\end{equation}
and
\begin{equation}\label{L-L}
    [L_1,L_2] = L_1 - L_2 \ .
\end{equation}
Define the time-dependent commutative generator
\begin{equation}\label{}
    L_t = \frac{\omega(t)}{2} L_0 + \frac{\delta(t)}{2} ( \mu_1 L_1 + \mu_2 L_2) + \frac{\gamma(t)}{2} L_z\ ,
\end{equation}
with $\mu_1,\mu_2 \geq 0$ and $\mu_1 + \mu_2=1$. Defining
\begin{equation}\label{}
    \Omega(t) = \int_0^t \omega(u)du\ ; \ \ \Delta(t) = \int_0^t \delta(u)du\ ; \ \ \Gamma(t) = \int_0^t \gamma(u)du\ ,
\end{equation}
one finds that $L_t$ is a legitimate generator iff $\Delta(t) \geq 0$ and $\Gamma(t) \geq 0$.
The following evolution of $\rho$ has the following form: the off-diagonal elements evolve according to
$$\rho_{12} \ra e^{\Omega(t) + \frac 12 \Delta(t) + \Gamma(t)} \rho_{12}\ , $$
and diagonal elements
\begin{eqnarray*}
  \rho_{11} &\ra & \rho_{11}\, e^{-\Delta(t)} + \mu_1 \Big[ 1 - e^{-\Delta(t)}  \Big] \ ,\\
  \rho_{22} &\ra & \rho_{22}\, e^{-\Delta(t)} + \mu_2 \Big[ 1 - e^{-\Delta(t)}  \Big] \ .
\end{eqnarray*}
If $\Delta(t) \ra \infty$ for $t\ra \infty$, then dynamics possesses an equilibrium state
$$  \rho_t \ \ \longrightarrow \ \ \left( \begin{array}{cc} \mu_1 & 0 \\ 0 & \mu_2 \end{array} \right) \ . $$

\end{Example}

\begin{Example}[Random unitary qubit dynamics] \label{EX-RU} Consider the following time-dependent generator
\begin{equation}\label{Pauli}
    L_t(\rho) = \frac 12 \sum_{k=1}^3 \gamma_k(t) (\sigma_k \rho\, \sigma_k - \rho)\ ,
\end{equation}
where $\{\sigma_1,\sigma_2,\sigma_3\}$ are Pauli matrices.  It is easy to prove that $[L_t,L_u]=0$ and hence $L_t$ generates a legitimate dynamical map iff
$$  \Gamma_1(t) \geq 0 \ ; \ \ \Gamma_2(t) \geq 0 \ ; \ \ \Gamma_3(t) \geq 0 \ , $$
where $\Gamma_k(t) = \int_0^t \gamma_k(u)du$.  One finds \cite{RUD} that the corresponding dynamical map $\Lambda_t$ is given by
\begin{equation}\label{RU}
    \Lambda_t(\rho) = \sum_{\alpha=0}^3 p_\alpha(t) \sigma_\alpha \rho\, \sigma_\alpha \ ,
\end{equation}
where $\sigma_0 = \mathbb{I}_2$ and
\begin{eqnarray*}
p_0(t) &=&\frac{1}{4}\, [1+ \lambda_3(t) + \lambda_2(t) + \lambda_1(t)] \ , \\
p_1(t) &=&\frac{1}{4}\, [1- \lambda_3(t) - \lambda_2(t) + \lambda_1(t)] \ ,\\
p_2(t) &=&\frac{1}{4}\, [1- \lambda_3(t) + \lambda_2(t) - \lambda_1(t)] \ ,\\
p_3(t) &=&\frac{1}{4}\, [1+ \lambda_3(t) - \lambda_2(t) - \lambda_1(t)] \ ,
\end{eqnarray*}
with
$$ \lambda_1(t) = e^{-\Gamma_2(t) - \Gamma_3(t)}\ , $$
and similarly for $\lambda_2(t)$ and $\lambda_3(t)$. Interestingly $\Lambda_t(\sigma_k)=\lambda_k(t)\sigma_k$.
The formula (\ref{RU}) defines so-called random unitary dynamics. Note that
$$ p_0(t) + p_1(t) + p_2(t) +  p_3(t)=1 \ . $$
Moreover, $p_\alpha(t) \geq 0$ for $\alpha=0,1,2,3$ iff $\Gamma_k(t) \geq 0$ for $k=1,2,3$. Note that $\Lambda_t$ is unital. Actually, in the case of qubit any unital dynamical map is random unitary, i.e.
\begin{equation}\label{}
    \Lambda_t(\rho) = \sum_k p_k(t) U_k(t) \rho U_k^\dagger(t)\ ,
\end{equation}
where $p_k(t)$ defines time-dependent probability distribution and $U_k(t)$ is a family of time-dependent unitary matrices. It is no longer true for $n$-level systems with $n>2$.
\end{Example}


\section{Markovian evolution --- divisible dynamical maps}   \label{SEC-C3}

We call a dynamical map $\Lambda_t$ {\em divisible} if for any $t \geq s \geq 0$ one has the following decomposition
\begin{equation}\label{}
    \Lambda_t = V_{t,s}\, \Lambda_s \ ,
\end{equation}
with a completely positive propagator $V_{t,s}$. Note, that if $\Lambda_t$ is invertible then
\begin{equation}\label{}
    V_{t,s} = \Lambda_t \, \Lambda_s^{-1} \ ,
\end{equation}
and hence  $V_{t,s}$ satisfies an inhomogeneous composition law
\begin{equation}\label{}
    V_{t,s} V_{s,u} = V_{t,u}\ ,
\end{equation}
for any $t\geq s \geq u$. The above formula provides a generalization of the semi-group composition low (\ref{c-low}). In this paper we accept the following definition of Markovian evolution: {\em a
dynamical map $\Lambda_t$ corresponds to Markovian evolution if and only if it is divisible}.
Interestingly, the property of being Markovian (or divisible) is fully characterized in terms of the local generator $L_t$. Note, that if $\Lambda_t$ satisfies (\ref{ME}) then $V_{t,s}$ satisfies
\begin{equation}\label{Local-V}
    \frac{d}{dt}\, V_{t,s} = L_t V_{t,s}\ ,     \ \ \ V_{s,s}=\oper\ ,
\end{equation}
and the corresponding solution reads
\begin{equation}\label{}
 V_{t,s} = {\rm T} \, \exp\left( \int_s^t L_u du \right)\ .
\end{equation}
It is clear that $\Lambda_t = V_{t,0}$ which shows that divisibility puts very strong requirements upon the dynamical map $\Lambda_t$.  One proves \cite{Alicki} the following

\begin{thm}
The map $\Lambda_t$ is divisible if and only if $L_t$ is a GKSL generator  for all $t$, that is,
\begin{equation}\label{}
    L_t(\rho) = -i[H(t),\rho] + \frac 12 \sum_k \left( [V_k(t),\rho V^\dagger_k(t)] + [V_k(t)\rho,V^\dagger_k(t)] \right) \ ,
\end{equation}
with time-dependent Hamiltonian $H(t)$ and noise operators $V_k(t)$.
\end{thm}

\begin{remark} If \begin{equation}\label{}
    L_t = \omega(t) L_0 + a_1(t) L_1 + \ldots + a_N(t) L_N\ ,
\end{equation}
where  $L_0(\rho) =  -i[H,\rho]$, and for $\alpha>0$ the generators $L_\alpha$ are purely dissipative and linearly independent, then
$L_t$ generates Markovian  evolution if and only if $a_1(t),\ldots,a_N(t) \geq 0$.
\end{remark}

\begin{Example} Consider a qubit generator
\begin{equation}\label{}
L_t  = -i[H(t),\,\cdot \, ] + \frac{\gamma_1(t)}{2}  L_1 + \frac{\gamma_2(t)}{2}  L_2 + \frac{\gamma(t)}{2} L_3\ ,
\end{equation}
where $\{L_1,L_2,L_3\}$ are defined in (\ref{LLL}). $L_t$ gives rise to Markovian evolution if and only if $\gamma_1(t),\gamma_2(t),\gamma(t) \geq 0$. We stress that due to the non-commutativity of $L_t$ it is not easy to find the corresponding solution defined in terms of the $T$-product formula (\ref{Dyson}) (cf. Section \ref{SEC-GEN}). In particular if we consider the special case corresponding to
\begin{equation}\label{}
    L_t(\rho) = \frac 12 {\gamma(t)} L_3(\rho) = \frac 12 {\gamma(t)} (\sigma_z \rho \sigma_z - \rho)\ ,
\end{equation}
and introduce $   \Gamma(t) = \int_0^t \gamma(\tau)d\tau$, then it is clear that
\begin{equation}\label{}
    \Lambda_t(\rho) = \frac 12 \Big[ 1+ e^{-\Gamma(t)} \Big]\, \rho + \frac 12 \Big[ 1- e^{-\Gamma(t)} \Big]\, \sigma_z \rho \sigma_z\ ,
\end{equation}
and hence
\begin{enumerate}

\item $L_t$ is a legitimate generator iff $\,\Gamma(t)\geq 0$,

\item $L_t$ generates Markovian evolution iff $\gamma(t) \geq 0$,

\item $L_t$ generates Markovian semigroup iff $\gamma(t) = {\rm const.} >0$.

\end{enumerate}

\end{Example}
Divisible maps possess several important properties \cite{JPB}. Note that Proposition \ref{TH-D} implies the following

\begin{pro} If $\Lambda_t : \Mn \ra \Mn$ is a dynamical map, then
\begin{eqnarray}\label{DFS-dt-X}
    \frac{d}{dt}\, ||[\oper_n \ot\Lambda_t](X)||_1 \leq 0\ ,
\end{eqnarray}
for any Hermitian operator $X \in \Mn \ot \Mn$.
\end{pro}
Proof: one has
\begin{eqnarray*}
  \frac{d}{dt}\, ||[\oper_n \ot\Lambda_t](X)||_1  &= & \lim_{\epsilon\rightarrow\, 0+} \frac{1}{\epsilon}\, \Big[
  ||[\oper_n \ot\Lambda_{t+\epsilon}](X)||_1 - ||[\oper_n \ot\Lambda_t](X)||_1 \Big] \\
  &= & \lim_{\epsilon\rightarrow\, 0+} \frac{1}{\epsilon}\, \Big[ ||[\oper_n \ot V_{t+\epsilon,t}\Lambda_{t}](X)||_1 - ||[\oper_n \ot\Lambda_t](X)||_1 \Big] \\
   &\leq&  \lim_{\epsilon\rightarrow\, 0+} \frac{1}{\epsilon}\, \Big[ ||[\oper_n \ot \Lambda_{t}](X)||_1 -  ||[\oper_n \ot \Lambda_{t}](X)||_1 \Big]=0\ ,
\end{eqnarray*}
where we have used the divisibility property $\Lambda_{t+\epsilon} = V_{t+\epsilon,t} \Lambda_t$ and Proposition \ref{TH-D} for the  map $\Phi=\oper_n \ot V_{t+\epsilon,t}$.  \hfill $\Box$

\begin{remark} Note that taking $X = \mathbb{I}_n \ot x$ one immediately gets
\begin{eqnarray}\label{}
    \frac{d}{dt}\, ||\Lambda_t(x)||_1 \leq 0\ ,
\end{eqnarray}
for any Hermitian operator $x\in \Mn$. In particular if $X=\rho - \sigma$ one arrives at
\begin{eqnarray}\label{BLP-X}
    \frac{d}{dt}\, ||\Lambda_t(\rho-\sigma)||_1 \leq 0\ ,
\end{eqnarray}
for any pair of density matrices $\rho$ and $\sigma$.
\end{remark}

\begin{remark}\label{R-6} Actually,  the formula (\ref{BLP-X}) was used in \cite{BLP} as a definition of Markovian dynamics: the dynamics $\Lambda_t$ is Markovian iff
\begin{equation}\label{M-BLP}
\frac{d}{dt}\, D[\Lambda_t(\rho),\Lambda_t(\sigma)] \leq 0\ ,
\end{equation}
for all initial states $\rho,\sigma\in \mathfrak{S}_n$. It is clear that if $\Lambda_t$ is divisible, then (\ref{M-BLP}) is satisfied. Note, however, that the converse needs not be true. Consider the following example: let the dynamics $\Lambda_t$ be governed by the local in time generator
\begin{equation}\label{Tr-omega}
    L_t(\rho) = \gamma(t) \left( \omega_t \, {\rm Tr}\, \rho - \rho \right) \ ,
\end{equation}
where $\omega_t$ is a family of Hermitian operators satisfying ${\rm Tr}\,\omega_t = 1$.
The above generator gives rise to Markovian evolution iff $L_t$ has the standard GKSL form \cite{GKS,Lindblad} for all $t\geq 0$, that is, iff $\gamma(t) \geq 0$ and $\omega_t$ defines a legitimate state, i.e. $\omega_t \geq 0$. The corresponding solution of the master equation $\dot{\rho}_t = L_t \rho_t$ with an initial condition $\rho$ reads as follows
  $$  {\rho}_t = e^{-\Gamma(t)}\widetilde{\rho}_t\ , $$
where as usual $\Gamma(t) = \int_0^t \gamma(u)du$ and $\widetilde{\rho}_t$ satisfies
$$\partial_t \widetilde{\rho}_t = \gamma(t) \omega_t {\rm Tr}\, \widetilde{\rho}_t\ . $$
One has
$$ \partial_t [{\rm Tr}\, \widetilde{\rho}_t] = \gamma(t) {\rm Tr}\, \widetilde{\rho}_t\ , $$
which implies
$  {\rm Tr}\, \widetilde{\rho}_t  = e^{\Gamma(t)} {\rm Tr}\,  \widetilde{\rho} = e^{\Gamma(t)} {\rm Tr}\, {\rho}$ due to $\Gamma(0)=1$.
Therefore, one arrives at the following equation
$$\partial_t \widetilde{\rho}_t = \gamma(t) e^{\Gamma(t)} \omega_t {\rm Tr}\,{\rho}\ , $$
with the corresponding solution
$$  \widetilde{\rho}_t = \widetilde{\rho} + \left[\int_0^t \gamma(u) e^{\Gamma(u)} \omega_u \, du\right] {\rm Tr}\,{\rho}\ . $$
Finally,
\begin{equation}\label{}
    \rho_t = \Lambda_t(\rho) = e^{-\Gamma(t)} \rho + [1- e^{-\Gamma(t)}] \Omega_t\, {\rm Tr}\rho\ ,
\end{equation}
where
$$\Omega_t = \frac{1}{e^{\Gamma(t)}-1} \int_0^t \gamma(\tau) e^{\Gamma(\tau)} \, \omega_\tau d\tau \ . $$
It is therefore clear
that $L_t$ generates a legitimate quantum evolution iff $\Gamma(t) \geq 0$ and $\Omega_t \geq 0$, that is, $\Omega_t$ defines a legitimate state (note, that ${\rm Tr}\, \Omega_t=1$). In particular, if $\omega_t = \omega$ is time-independent, then $\Omega_t = \omega$ and the solution simplifies to  a convex combination of the initial state $\rho$ and the asymptotic invariant state $\omega$: $\rho_t = e^{-\Gamma(t)} \rho + [1- e^{-\Gamma(t)}]\, \omega$. One easily shows that the evolution is Markovian iff $\gamma(t) \geq 0$ and $\omega_t$ is a legitimate density operator. Consider now  condition (\ref{M-BLP}). One has $\rho_t - \sigma_t = e^{-\Gamma(t)} (\rho-\sigma)$ and hence
\begin{equation}
    \frac{d}{dt}\, ||\rho_t - \sigma_t||_1  = - \gamma(t)\, e^{-\Gamma(t)} ||\rho-\sigma||_1 \leq 0 \ , \nonumber
\end{equation}
implies only $\gamma(t) \geq 0$ but says nothing about $\omega_t$. It shows that any $\omega_t$ which gives rise to $\Omega_t \geq 0$ leads to the evolution satisfying condition (\ref{M-BLP}) but only $\omega_t \geq 0$ gives rise to Markovian dynamics. Hence, we may have non-Markovian  dynamics  which satisfies condition (\ref{M-BLP}) for all $t\geq 0$. In this case $\omega_t$, contrary to $\Omega_t$, is no longer a state.  The interested reader will find other examples showing that (\ref{M-BLP}) may differ from divisibility property in \cite{inne1,inne2,inne3}.

\end{remark}

\section{The general structure of local generators}  \label{SEC-GEN}

Consider now a dynamical map  $\Lambda_t$  and suppose that it can be represented as follows
\begin{equation}\label{Z}
    \Lambda_t = e^{Z_t}  \ ,
\end{equation}
where $Z_t : \Mn \ra \Mn$ is a legitimate GKSL generator for all $t \geq 0$ and $Z_0 = 0$ which guarantees that $\Lambda_0 = \oper_n$. It is clear that such representation enforces $\Lambda_t$ to be a family of quantum channels for $t\geq 0$. Let us assume that $Z_t$ is differentiable and let
\begin{equation}\label{}
    Z_t = \int_0^t X_\tau d\tau\ .
\end{equation}
Assume now that $\Lambda_t$ satisfies local in time Master Equation (\ref{ME}). The corresponding local generator is given by
\begin{equation}\label{}
    L_t = \dot{\Lambda}_t \Lambda_t^{-1} \ ,
\end{equation}
provided that the inverse map $\Lambda_t^{-1}$ does exist. Note, however, that the representation (\ref{Z}) implies that $\Lambda_t^{-1} = e^{-Z_t}$ and hence $L_t$ is well defined. To find $L_t$ one needs to compute $\partial_t\Lambda_t$. In order to do it we use the following Wilcox formula \cite{Wilcox}
\begin{eqnarray}\label{}
    \partial_t\, e^{Z_t} &=& \int_0^1 e^{s Z_t}\, \dot{Z}_t\, e^{(1-s)Z_t}\, ds \ ,
\end{eqnarray}
and hence
\begin{equation}\label{L-GEN}
    L_t = \dot{\Lambda}_t \Lambda_t^{-1} = \int_0^1 e^{s Z_t}\, X_t\, e^{-sZ_t}\, ds \ ,
\end{equation}
where we have used $X_t = \dot{Z}_t$. The formula (\ref{L-GEN}) establishes very general form of the local generator. The
construction of a legitimate generator is pretty simple, nevertheless, the formula (\ref{L-GEN}) is
highly nontrivial and the computation of $L_t$ out of
$Z_t$ might be highly complicated. This is the price we pay for
the simple representation of evolution  (\ref{Z}). Hence, we
have a kind of complementarity: either one uses T-product formula
(\ref{Dyson}) with a relatively simple generator or one avoids
the T-product through (\ref{Z}) but uses highly nontrivial generator
(\ref{L-GEN}). The advantage of our approach is that one knows how to
construct the generator (in practice it might be complicated) giving
rise to the legitimate quantum dynamics.

\begin{remark} Note, that if $Z_t$ is a commutative family, then
\begin{equation}\label{}
    0 = \partial_t [Z_t,Z_u] = [X_t,Z_u] \ ,
\end{equation}
and hence $e^{sZ_t}$ and $X_t$ commute. In this case formula (\ref{L-GEN}) simplifies to
\begin{equation}\label{}
    L_t = \int_0^1 e^{s Z_t}\, X_t\, e^{-sZ_t}\, ds \  = \int_0^1 X_t \, ds = X_t\ ,
\end{equation}
which shows that $L_t$ coincides with $X_t$. In the
noncommutative case this simple relation is no longer true.
\end{remark}

\begin{remark} It should be clear why we use the special representation $\Lambda_t = e^{Z_t}$ for the dynamical map. Using for example the Kraus representation
$$    \Lambda_t(\rho) = \sum_\alpha K_\alpha(t) \rho \, K_\alpha^\dagger(t)\ , $$
there is no simple way to calculate the inverse $\Lambda_t^{-1}$. If $\Lambda_t = e^{Z_t}$ we have the inverse $\Lambda_t^{-1} = e^{-Z_t}$ for free!
\end{remark}

To illustrate this approach let us consider the following instructive example of a qubit dynamics: let
\begin{equation}\label{}
    X_t = a_1(t) L_1 + a_2(t) L_2 \ ,
\end{equation}
where $L_1,L_2$ are defined in (\ref{LLL}) and $a_1(t),a_2(t)$ are real functions of time. One has therefore
\begin{equation}\label{}
    Z_t =  A_1(t) L_1 + A_2(t) L_2 \ ,
\end{equation}
where
\begin{equation}\label{}
    A_1(t) = \int_0^t a_1(u)\, du\ ; \ \ \ \   A_2(t) = \int_0^t a_2(u)\, du\ .
\end{equation}
It is clear that $\Lambda_t = e^{Z_t}$ defines a legitimate qubit dynamics if and only if $A_k(t) \geq 0$ for $k=1,2$. Our goal now is to find the corresponding local generator $L_t$ using our basic formula (\ref{L-GEN}). Note that a family $X_t : \Mq \ra \Mq$ provides a noncommutative family of maps due to the following commutation relation (cf. formula (\ref{L-L}))
\begin{equation}\label{L12}
    [L_1,L_2]= L_1 - L_2\ .
\end{equation}
Hence, $L_t$ is different from $X_t$. To compute $L_t$ via Wilcox formula (\ref{L-GEN}) let us observe that $L_1$ and $L_2$ close a Lie algebra and hence we may use well-known Lie algebraic methods. One has  the following quite involved expression for $L_t$
\begin{equation}\label{L.}
    L_t = \int_0^t e^{s[A_1(t) L_1 + A_2(t)L_2]} [ a_1(t) L_1 + a_2(t)L_2] e^{-s[A_1(t) L_1 + A_2(t)L_2]}\, ds\ .
\end{equation}
To deal with it we shall use  the well-known Baker-Campbell-Hausdorff (BCH) formula
\begin{equation}\label{}
    e^X Y e^{-X} = Y + [X,Y] + \frac{1}{2!} [X,[X,Y]] + \ldots  = \sum_{k=0}^\infty \frac{1}{k!}\, {\rm ad}_X^k Y\ ,
\end{equation}
where ${\rm ad}_X Y := [X,Y]$.  Using the simple commutation rule (\ref{L12}) one easily proves for $k\geq 1$
\begin{eqnarray*}
  {\rm ad}_Z^k L_1  &=& (-1)^k A^{k-1} A_2 (L_2-L_1) \ , \\
   {\rm ad}_Z^k L_2  &=& (-1)^k A^{k-1} A_1 (L_1-L_2) \ ,
\end{eqnarray*}
where we introduced
$$A=A_1 + A_2\ . $$
 Therefore, one gets
\begin{eqnarray*}
  && e^{sZ} L_1 e^{-sZ}  = \sum_{k=0}^\infty \frac{1}{k!}\, {\rm ad}_Z^k L_1 = L_1 +  \sum_{k=1}^\infty \frac{1}{k!}(-s)^k A^{k-1} A_2 (L_2-L_1) \\ &=& L_1 + (L_2-L_1) \frac{A_2}{A}  \sum_{k=1}^\infty \frac{1}{k!}(-s A)^k = L_1 + \frac{A_2}{A}(1-e^{-sA}) (L_2-L_1) \ ,
\end{eqnarray*}
and similarly
\begin{eqnarray*}
  && e^{sZ} L_2 e^{-sZ}  = L_2 + \frac{A_1}{A}(1-e^{-sA}) (L_1-L_2) \ ,
\end{eqnarray*}
where to simplify the notation we omit the time dependence of $A_1$ and $A_2$. Inserting
\begin{eqnarray}\label{ZL1Z}
    e^{sZ} L_1  e^{-sZ} =   \left( 1 - \frac{A_2}{A}\big[1-e^{-sA}\big] \right) L_1 + \frac{A_2}{A}[1-e^{-sA}] L_2\ ,
\end{eqnarray}
and
\begin{equation}\label{ZL2Z}
    e^{sZ} L_2  e^{-sZ} = \left( 1 - \frac{A_1}{A}[1-e^{-sA}] \right) L_2 + \frac{A_1}{A}[1-e^{-sA}] L_1\ ,
\end{equation}
into (\ref{L.}) one obtains finally
\begin{eqnarray}\label{}
    L_t = b_1(t) L_1 + b_2(t) L_2\ ,
\end{eqnarray}
where the functions $b_1(t)$ and $b_2(t)$ are defined by
\begin{eqnarray}
  b_1 &=& a_1 -f \ , \\
  b_2  &=& a_2 + f \ ,
\end{eqnarray}
and the function $f(t)$ reads
\begin{equation}\label{}
    f = e^{-A} \frac{W}{A}\ ,
\end{equation}
with $W = a_1 A_2 - a_2 A_1$ being the Wronskian of a pair $\{A_1(t),A_2(t)\}$. Hence, the local generator $L_t$ has exactly the same structure as $X_t$ but with functions $a_k(t)$ replaced by $b_k(t)$. Note, that if $a_1(t)$ and $a_2(t)$ are not linearly independent, i.e. $a_2(t) = \lambda a_1(t)$, then the Wronskian vanishes and $b_k=a_k$ which implies $L_t=X_t$. This result should be clear since  in this case $X_t$ defines a commutative family $X_t = a_1(t)[L_1 + \lambda L_2]$.

To summarize: the local generator $L_t$ is related with $X_t$ by the following simple relation
\begin{equation}\label{L-X}
    L_t = X_t - f(t)[L_1 - L_2]\ .
\end{equation}
Roughly speaking the function $f(t)$ measures the non-commutativity of the dynamical map $\Lambda_t$. Note that we have two representations of $\Lambda_t$: as a solution of the Master Equation $\Lambda_t = {\rm T} \exp\left( \int_0^t L_u du\right)$ and $\Lambda_t = \exp(\int_0^t X_u du)$. Using the formula (\ref{L-X}) one arrives at the following interesting observation
\begin{equation}\label{}
    {\rm T} \exp\left( \int_0^t ( X_u - f(u)[L_1-L_2]) du\right) = \exp\left(\int_0^t X_u du \right)\ ,
\end{equation}
which shows that the role of T-product is just to kill the unwanted term $f(u)[L_1-L_2]$!

Note, that the representation $\Lambda_t = \exp(\int_0^t X_u du)$ is legitimate if and only if $A_1(t),A_2(t) \geq 0$. However, this condition says nothing about positivity of $b_1(t)$ and $b_2(t)$. Even the integrals
\begin{equation}\label{BBF}
    B_1 = A_1 - F \ ; \ \ \ B_2 = A_2 + F\ ,
\end{equation}
with $F(t) = \int_0^t f(u)du$ need not be positive. Only the sum
\begin{equation}\label{}
    B_1(t) + B_2(t) = A_1(t) + A_2(t) \geq 0\ ,
\end{equation}
is fully controlled by $A_1$ and $A_2$. Interestingly, one can prove the following relations between $\{a_1,a_2\}$ and $\{b_1,b_2\}$:

\begin{pro} If $\, b_1(t),b_2(t) \geq 0$, then $\, A_1(t),A_2(t) \geq 0$.
\end{pro}
This statement is obvious: if $\, b_1(t),b_2(t) \geq 0$, then $L_t$ generates divisible dynamical map and hence $\Lambda_t = e^{Z_t}$ implies that $Z_t$ is a legitimate GKSL generator for $t\geq 0$ which in turn is equivalent to  $\, A_1(t),A_2(t) \geq 0$. One has also the {\em dual} result

\begin{pro} If $\, a_1(t),a_2(t) \geq 0$, then $\, B_1(t),B_2(t) \geq 0$.
\end{pro}
The proof is very easy: due to (\ref{BBF}) it is enough to show that
\begin{equation}\label{}
    - A_2(t) \leq F(t)  \leq A_1(t)\ ,
\end{equation}
for all $t\geq 0$. One has
\begin{eqnarray*}\label{}
    F(t) &=& \int_0^t du\, e^{-A(u)} \Big[ a_1(u) A_2(u) - a_2(u) A_1(u) \Big]/A(u) \\
    &=& \int_0^t du\, \Big[ e^{-A(u)} a_1(u)  - e^{-A(u)} A_1(u)[a_1(u) + a_2(u)]/A(u) \Big] \\
    &\leq & \int_0^t du\, e^{-A(u)} a_1(u) \leq \int_0^t du\, a_1(u) = A_1(t) \ ,
\end{eqnarray*}
and in the same way one proves that $-F(t) \leq A_2(t)$.

\begin{remark} Actually, the Master Equation with $L_t = b_1(t) L_1 + b_2(t) L_2$  may be easily solved. Defining
\begin{equation}\label{}
  \omega_t = p_1(t) |1\>\<1| + p_2(t) |2\>\<2|\ ,
\end{equation}
with $p_1(t) = b_2(t)/b(t)$ and $p_2(t) = b_1(t)/b(t)$ one easily finds
\begin{equation}\label{}
b_1(t) L_1 + b_2(t) L_2 =  b(t) [\omega_t {\rm Tr}\rho - \rho ] - \frac 14 b(t) L_z\ .
\end{equation}
Note that generator $b(t) [\omega_t {\rm Tr}\rho - \rho ]$ was already considered in Remark \ref{R-6} (see formula (\ref{Tr-omega}). Moreover, one easily checks that this generator commutes with $L_z$ and hence one finds the following formula for $\Lambda_t$
\begin{eqnarray*}
  \Lambda_t(\rho)  &=& \frac 12 e^{-B(t)} \Big[ (1 + e^{B(t)/2})\rho + (1 - e^{B(t)/2})\sigma_z\rho\sigma_z \Big] + (1-e^{-B(t)}) \Omega_t\ , 
\end{eqnarray*}
where 
$$\Omega_t = \frac{1}{e^{B(t)}-1} \int_0^t b(\tau) e^{B(\tau)} \, \Big[p_1(\tau)|1\>\<1| + p_2(\tau)|2\>\<2| \Big] d\tau \ . $$
To check for complete positivity of $\Lambda_t$ one computes $[\oper_2 \ot \Lambda_t](P^+_2)$:
\begin{equation*}\label{}
  \left( \begin{array}{cc|cc} e^{-B} + (1-e^{-B})\Omega_{11} & \cdot & \cdot & e^{-B/2} \\
  \cdot & (1-e^{-B})\Omega_{22} &  \cdot & \cdot \\ \hline
  \cdot & \cdot & (1-e^{-B})\Omega_{11} & \cdot \\
  e^{-B/2} & \cdot & \cdot & e^{-B} + (1-e^{-B})\Omega_{22} \end{array} \right)\ ,
\end{equation*}
where we skiped the time dependence. Now, $[\oper_2 \ot \Lambda_t](P^+_2)\geq 0$ iff $B(t) \geq 0$ and $\Omega_t \geq 0$. The last condition is equivalent to
\begin{equation}\label{}
  0 \leq \int_0^t b_k(\tau) e^{B(\tau)} d\tau \leq e^{B(t)} - 1 \ ,
\end{equation}
for $k=1,2$. This way we derived conditions for $b_1(t)$ and $b_2(t)$. Note, however, that in order to do this we have to solve the original Master Equation! Only the condition $B(t) \geq 0$ was obtained via the Lie algebraic method. 
\end{remark}

\begin{cor} If $X_t$ is GKSL generator, then $\int_0^t L_u du$ is GKSL generator.
If $L_t$ is GKSL generator (the dynamics is Markovian), then $\int_0^t X_u du$ is GKSL generator.
\end{cor}

\begin{pro}
The corresponding solution $e^{Z_t}$ may be represented as follows
\begin{equation}\label{Z-nu12}
    e^{Z_t} = e^{\nu_1(t)\, L_1} e^{\nu_2(t)\, L_2}\ ,
\end{equation}
where
\begin{equation}\label{nu-12}
    \nu_1 = \ln \left( \frac{A}{A_1 e^{-A} + A_2}\right) \ , \ \ \ \nu_2 = \ln\left( \frac{A_1 + A_2 e^A}{A}\right)\ ,
\end{equation}
and hence $\nu_1 + \nu_2 = A$.
\end{pro}
Proof: Since $\{L_1,L_2\}$ form a Lie algebra one has
\begin{equation}\label{}
    e^{s(A_1 L_1 + A_2 L_2)} = e^{u_1(s)L_1}   e^{u_2(s)L_2}\ ,
\end{equation}
where $u_1(s),u_2(s)$ are real functions of a parameter $s$ satisfying $u_1(0)=u_2(0)=0$. Differentiating both sides with respect to $s$ one gets
\begin{eqnarray*}\label{}
   && e^{s(A_1 L_1 + A_2 L_2)}(A_1 L_1 + A_2 L_2)  \\
   &=& u'_1(s) L_1 e^{u_1(s)L_1}   e^{u_2(s)L_2} +  u'_2(s)e^{u_1(s)L_1}   e^{u_2(s)L_2}  L_2 \\
   &=& u'_1(s) L_1 e^{s(A_1 L_1 + A_2 L_2)}  +  u'_2(s)e^{s(A_1 L_1 + A_2 L_2)} L_2 \ ,
\end{eqnarray*}
where $u'(s) = du(s)/ds$. Multiplying both sides by $e^{-s(A_1 L_1 + A_2 L_2)}$ leads to
\begin{eqnarray*}\label{}
   A_1 L_1 + A_2 L_2 = u'_1(s) L_1 + u'_2(s) e^{s(A_1 L_1 + A_2 L_2)} L_2 e^{-s(A_1 L_1 + A_2 L_2)}\ ,
\end{eqnarray*}
and hence using (\ref{ZL2Z}) one obtains the following equations for unknown functions $u_1$ and $u_2$:
\begin{eqnarray}
  A_1 &=& u'_1(s) \Big[ 1 - \frac{A_2}{A} (1 - e^{s A}) \Big] \ ,  \\
  A_2 &=& u'_1(s) \frac{A_2}{A} (1 - e^{s A}) + u'_2(s) \ .
\end{eqnarray}
Using
\begin{equation}\label{}
    \int_0^1 \frac{dx}{1+ae^{b x}} = \frac 1b \ln \frac{1+a}{e^{-b} + a}\ ,
\end{equation}
one finds
\begin{equation}\label{}
    u_1(1) = A \int_0^1 \frac{ds}{1+ (A_2/A_1) e^{sA}} = \ln \left(\frac{A}{A_1 e^{-A} + A_2}\right) \ ,
\end{equation}
and
\begin{equation}\label{}
    u_2(1) = \ln\left( \frac{A_1 + A_2 e^{A}}{A} \right)\ .
\end{equation}
Finally, one takes $\nu_1 := u_1(1)$ and   $\nu_2 := u_2(1)$. It is clear that  $\nu_1,\nu_2 \geq 0$ which proves that $e^{\nu_1(t)\, L_1}$ and $e^{\nu_2(t)\, L_2}$ are CPTP maps. \hfill $\Box$

\begin{remark} This example may be easily generalized: suppose that $\{L_1,\ldots,L_N\}$ is a set of GKSL generators closing a Lie algebra
\begin{equation}\label{}
    [L_i,L_j] = \sum_{k=1}^N c_{ij}^k L_k\ ,
\end{equation}
and consider
\begin{equation}\label{}
    X_t = a_1(t) L_1 + \ldots + a_N(t) L_N\ ,
\end{equation}
such that
\begin{equation}\label{}
    A_1(t) = \int_0^t a_1(u)du \geq 0 \ ; \ \ \ldots \ \ ;  A_N(t) = \int_0^t a_N(u)du \geq 0\ .
\end{equation}
It should be clear that Wilcox formula leads to
\begin{equation}\label{}
    L_t = b_1(t) L_1 + \ldots + b_N(t) L_N\ ,
\end{equation}
where the functions $\{b_1(t),\ldots,b_N(t)\}$ are fully determined by $\{a_1(t),\ldots,a_N(t)\}$ and the structure constants $c_{ij}^k$ of the above Lie algebra. There are well-developed algebraic methods (like Wei-Norman or Magnus expansions \cite{Magnus,Wei} see also the recent review \cite{M-review}) which may help to deal with such dynamical problems.
\end{remark}

\section{Conclusions}

We provided a basic introduction to the time-local description of open quantum systems. The full characterization of Markovian semigroups, commutative evolutions and dynamics corresponding to divisible maps (Markovian evolution) is provided. In the non-commutative case the Wilcox formula shows the intricate structure of time-local generator $L_t$ and partially explains the problem of finding necessary and sufficient conditions which do guarantee a legitimate quantum evolution generated by $L_t$. Recently, there is an increasing interest in the non-Markovian quantum evolutions and quantum memory effects. For recent papers devoted to both theoretical and experimental aspects of quantum evolution with memory see e.g. a collection of papers in  \cite{REV} and references therein. See also Haikka and Maniscalco \cite{Haikka} in this volume.





\section*{Acknowledgments}

This work was partially supported by the National Science Center project
DEC-2011/03/B/ST2/00136. Many thanks to Andrzej Kossakowski for sharing his passion for open quantum systems.

\end{document}